\documentstyle[12pt]{article}
\newlength{\extraspace}
\newlength{\extraspaces}
\newcommand{\be}{\begin{equation}
\addtolength{\abovedisplayskip}{\extraspaces}
\addtolength{\belowdisplayskip}{\extraspaces}
\addtolength{\abovedisplayshortskip}{\extraspace}
\addtolength{\belowdisplayshortskip}{\extraspace}}
\newcommand{\ee}{\end{equation}}

\newcommand{\ba}{\begin{eqnarray}
\addtolength{\abovedisplayskip}{\extraspaces}
\addtolength{\belowdisplayskip}{\extraspaces}
\addtolength{\abovedisplayshortskip}{\extraspace}
\addtolength{\belowdisplayshortskip}{\extraspace}}
\newcommand{\ea}{\end{eqnarray}}

\newcommand{\newsection}[1]{
\vspace{15mm}
\pagebreak[3]
\addtocounter{section}{1}
\setcounter{equation}{0}
\setcounter{subsection}{0}
\setcounter{footnote}{0}
\begin{flushleft}
{\large\bf \thesection. #1}
\end{flushleft}
\nopagebreak
\medskip
\nopagebreak}

\newcommand{\Tr}{{\rm Tr}}
\newcommand{\Dmrns}{{\cal D}_{\mu\rho,\nu\sigma}}

\begin{document}

\addtolength{\baselineskip}{.8mm}

{\thispagestyle{empty}

\noindent \hspace{1cm}  \hfill May 97 \hspace{1cm}\\
\mbox{}                 \hfill IFUP--TH 19/97 \hspace{1cm}\\

\begin{center}
\vspace*{1.0cm}
{\large\bf Field strength correlators in full QCD
 \footnote{Partially supported by MURST (Italian Ministry of the University 
 and of Scientific and Technological Research) and by the EC contract 
 CHEX--CT92--0051.} }\\
\vspace*{1.0cm}
{\large M. D'Elia, A. Di Giacomo, E. Meggiolaro
  \footnote{Postal address: Enrico Meggiolaro, Dipartimento di Fisica, 
  Universit\`a di Pisa, Piazza Torricelli 2, I--56100 Pisa, Italy.
  Fax nr.: +39--50--911313. 
  E--mail address: enrico.meggiolaro@sns.it .} 
}\\
\vspace*{0.5cm}{\normalsize
{Dipartimento di Fisica, \\
Universit\`a di Pisa, \\
and INFN, Sezione di Pisa,\\
I--56100 Pisa, Italy.}}\\
\vspace*{2cm}{\large \bf Abstract}
\end{center}
\noindent
We study, by numerical simulations on a lattice, the behaviour of the
gauge--invariant two--point correlation functions of the gauge field 
strengths in the QCD vacuum with dynamical fermions.\\
\vspace{1.0cm}
\noindent
(PACS code: 12.38.Gc)
}
\vfill\eject

\newsection{Introduction}

\noindent
A relevant role in hadron physics is played by the gauge--invariant two--point
correlators of the field strengths in the QCD vacuum. They are defined as
\be
\Dmrns(x) = \langle 0| 
\Tr \left\{ G_{\mu\rho}(x) S(x,0) G_{\nu\sigma}(0) S^\dagger(x,0) \right\}
|0\rangle ~,
\ee
where $G_{\mu\rho} = gT^aG^a_{\mu\rho}$ is the field--strength tensor and
$S(x,0)$ is the Schwinger phase operator needed to 
parallel--transport the tensor $G_{\nu\sigma}(0)$ to the point $x$.

They govern the effect of the gluon condensate on the level 
splittings in the spectrum of heavy $Q \bar{Q}$ bound states
\cite{Gromes82,Campostrini86,Simonov95}.
They are the basic quantities in models of stochastic confinement of colour
\cite{Dosch87,Dosch88,Simonov89}
and in the description of high--energy hadron scattering
\cite{Nachtmann84,Landshoff87,Kramer90,Dosch94}.

These correlators have been determined on the lattice in the
{\it quenched} (i.e., pure--gauge) theory, with gauge group $SU(2)$
\cite{Campostrini84}, and also in the {\it quenched} $SU(3)$ theory
in the range of physical distances between 0.1 and 1 fm
\cite{DiGiacomo92,myref97}.
In this paper we compute them in {\it full} QCD, i.e., we 
also include the effects of dynamical fermions.

The technique used is the same as in Refs. \cite{DiGiacomo92,myref97}.
The basic idea  is to remove the effects of short--range fluctuations on 
large distance correlators by a local {\it cooling} procedure
\cite{Campostrini89,DiGiacomo90}.
Freezing the links of QCD configurations one after the other, damps very 
rapidly the modes of short wavelength, but requires a number $n$ of cooling 
steps proportional to the square of the distance $d$ in lattice units to 
affect modes of wavelength $d$:
\be
n \simeq k d^2 ~.
\ee
Cooling is a kind of diffusion process.
If $d$ is sufficiently large, there will be a range of values of $n$ in 
which lattice artefacts due to short--range fluctuations have been removed, 
without touching the physics at distance $d$.
This removal will show up as a plateau in the dependence 
of the correlators on $n$. 

The results are presented in Sect. 2.
The determination was done at  $\beta = 5.35$ ($\beta = 6/g^2$, where $g$ 
is the coupling constant) on a $16^3 \times 24$ lattice 
with four flavours of {\it staggered} fermions and the Wilson action for the 
pure--gauge sector.
We have used a standard hybrid Monte Carlo (HMC) algorithm, in particular the 
so--called $\Phi$ algorithm described in detail in Ref. \cite{Gottlieb87}:
the trajectory length $\tau$ was taken to be 0.3 with a molecular--dynamics 
step size $\delta\tau = 0.004$.
The bare quark mass was chosen to be $a \cdot m_q = 0.01$ ($a$ being the 
lattice spacing), which should be a reasonable approximation to the chiral 
limit. 
A determination was also made for $a \cdot m_q = 0.02$, which we shall comment 
in the following. In Sect. 3 we discuss our results and give some concluding 
remarks.

\newsection{Computations and results}

\noindent
The parametrization of the correlators is taken from Refs.
\cite{Dosch87,Dosch88,Simonov89}:
\ba
\lefteqn{
\Dmrns(x) = (\delta_{\mu\nu}\delta_{\rho\sigma} - 
\delta_{\mu\sigma}\delta_{\rho\nu})
\left[ {\cal D}(x^2) + {\cal D}_1(x^2) \right] } \nonumber \\
& & + (x_\mu x_\nu \delta_{\rho\sigma} - x_\mu x_\sigma \delta_{\rho\nu} 
+ x_\rho x_\sigma \delta_{\mu\nu} - x_\rho x_\nu \delta_{\mu\sigma})
{\partial{\cal D}_1(x^2) \over \partial x^2} ~.
\ea
${\cal D}$ and ${\cal D}_1$ are invariant functions of $x^2$. We work in 
the Euclidean theory.

It is convenient to define a ${\cal D}_\parallel(x^2)$ and a 
${\cal D}_\perp(x^2)$ as follows:
\ba
{\cal D}_\parallel &\equiv& {\cal D} + {\cal D}_1 + x^2 {\partial{\cal D}_1
\over \partial x^2} ~, \nonumber \\
{\cal D}_\perp &\equiv& {\cal D} + {\cal D}_1 ~.
\ea
On the lattice we can define a lattice operator $\Dmrns^L$, which is
proportional to $\Dmrns$ in the na\"\i ve continuum limit, i.e., when the 
lattice spacing $a \to 0$ \cite{DiGiacomo92,myref97}.
Making use of the definition (2.2) we can thus write, in the same limit,
\ba
{\cal D}_\parallel^L(\hat d a) \mathop\sim_{a\to0} a^4 
{\cal D}_\parallel(d^2 a^2) + {\cal O}(a^6) ~,\nonumber \\
{\cal D}_\perp^L(\hat d a) \mathop\sim_{a\to0} a^4 
{\cal D}_\perp(d^2 a^2) + {\cal O}(a^6) ~.
\ea
Higher orders in $a$ in Eq. (2.3) as well as possible multiplicative 
renormalizations are removed by cooling the quantum fluctuations at the scale 
of the lattice spacing, as explained in the Introduction. 

The only scale in our system is the lattice spacing $a$: its value in 
physical units depends on $\beta$.
Here we work with only one value of $\beta$, so we could present our 
results directly in units of $a$. However, in order to facilitate the 
comparison with our previous works \cite{DiGiacomo92,myref97} we shall use 
the familiar parametrization
\be
a(\beta) = {1\over\Lambda_F} f(\beta) ~,
\ee
where the {\it scaling function} $f(\beta)$ is given by the usual two--loop 
expression:
\be
f(\beta) = \left({8\over25}\,\pi^2\beta\right)
^{ 231/625 } \exp\left(-{4\over25}\pi^2\beta\right) ~,
\ee
for gauge group $SU(3)$ and $N_f = 4$ flavours of quarks.
$\Lambda_F$ in Eq. (2.4) is an effective $\Lambda$--parameter for QCD in the 
lattice renormalization scheme, with $N_f = 4$ flavours of quarks.
With this parametrization:
\ba
{\cal D}_\parallel^L f(\beta)^{-4} &=& {1\over\Lambda_F^4}
{\cal D}_\parallel\left({d^2\over\Lambda_F^2}f^2(\beta)\right) ~,
\nonumber \\
{\cal D}_\perp^L f(\beta)^{-4} &=& {1\over\Lambda_F^4}
{\cal D}_\perp\left({d^2\over\Lambda_F^2}f^2(\beta)\right) ~.
\ea
We have measured the correlations on a $16^3 \times 24$ lattice at distances 
$d$ ranging from 3 to 8 lattice spacings and at $\beta = 5.35$.
At this value of $\beta$ the lattice spacing $a(\beta)$, extracted from the 
string tension or the $\rho$ mass, is of the order of $0.11$ fm  
\cite{Laermann93,Laermann94}, so that the lattice size is approximately 2 fm 
and therefore safe from infrared artefacts.
In Fig. 1 we display the results for ${\cal D}_\parallel^L f(\beta)^{-4}$ 
and ${\cal D}_\perp^L f(\beta)^{-4}$ versus $d_{\rm phys} = 
(d/\Lambda_F)$  $f(\beta)$, for a quark mass $a \cdot m_q = 0.01$.
Measurements have been done on a sample of 150 configurations, each separated 
by 15 HMC trajectories. Statistical errors have been estimated by using
a standard blocking procedure.
As in Ref. \cite{myref97} we have tried a best fit to these data with the 
functions
\ba
{\cal D} (x^2) &=& A_0 \exp\left( -|x|/\lambda_A \right) 
+ {a_0 \over |x|^4} \exp\left( -|x|/\lambda_a \right) ~,
\nonumber \\
{\cal D}_1 (x^2) &=& A_1 \exp\left( -|x|/\lambda_A \right)
+ {a_1 \over |x|^4} \exp\left( -|x|/\lambda_a \right) ~.
\ea
We have obtained the following results:
\ba
{A_0 \over \Lambda_F^4} = (1.74 \pm 0.24) \times 10^{10} &~,~&
{A_1 \over \Lambda_F^4} = (0.20 \pm 0.10) \times 10^{10} ~, \nonumber \\
a_0 = 0.71 \pm 0.03 &~,~& a_1 = 0.45 \pm 0.03 ~, \nonumber \\
{1\over\lambda_A\Lambda_F} = 544 \pm 27 &~,~&
{1\over\lambda_a\Lambda_F} = 42 \pm 11 ~,
\ea
with $\chi^2/N_{\rm d.o.f.} \simeq 0.5$. The continuum lines in Fig. 1
correspond to the central values of this best fit.

The corresponding results for the quark mass $a \cdot m_q = 0.02$ are 
displayed in Fig. 2, using for $\Lambda_F$ the same value adopted for
$a \cdot m_q = 0.01$: we shall comment on this point in the next Section.
Measurements have been done on a sample of 30 configurations, each separated 
by 20 HMC trajectories. A best fit to these data with the same functions (2.7)
gives the results
\ba
{A_0 \over \Lambda_F^4} = (3.48 \pm 0.42) \times 10^{10} &~,~&
{A_1 \over \Lambda_F^4} = (0.46 \pm 0.21) \times 10^{10} ~, \nonumber \\
a_0 = 0.66 \pm 0.03 &~,~& a_1 = 0.39 \pm 0.03 ~, \nonumber \\
{1\over\lambda_A\Lambda_F} = 631 \pm 23 &~,~&
{1\over\lambda_a\Lambda_F} = 61 \pm 20 ~,
\ea
with $\chi^2/N_{\rm d.o.f.} \simeq 0.7$. Again, the continuum lines in Fig. 2
correspond to the central values of this best fit.

\newsection{Discussion}

\noindent
Two quantities of physical interest can be extracted from our lattice 
determinations:
\begin{itemize}
\item[1)] the correlation length $\lambda_A$ of the gluon field strengths, 
defined in Eq. (2.7);
\item[2)] the so--called {\it gluon condensate}, defined as
\be
G_2 \equiv \langle {\alpha_s \over \pi} :G^a_{\mu\nu} G^a_{\mu\nu}: \rangle
~~~~~~~~~ (\alpha_s = {g^2 \over 4\pi}) ~.
\ee
\end{itemize}
Both of them play an important role in phenomenology. The correlation length
$\lambda_A$ is relevant for the description of vacuum models
\cite{Dosch87,Dosch88,Simonov89}.
The relevance of the gluon condensate was first pointed out by Shifman, 
Vainshtein and Zakharov (SVZ) \cite{SVZ79}. 
It is a fundamental quantity in QCD, in the context of the SVZ sum rules.

From lattice we extract $\lambda_A$ in units of lattice spacing $a$. To 
convert these units to physical units, the scale must be set by comparison 
with some physical quantity. This was done in Refs. 
\cite{Laermann93,Laermann94} by computing the string tension and the $\rho$ 
mass on the lattice and comparing them with the physical values.
Their lattice was identical to ours ($16^3 \times 24$): they also used the 
same value of $\beta$ ($5.35$), as well as $N_f = 4$ flavours of staggered 
fermions with mass $a \cdot m_q = 0.01$, as we did.
Their estimate for the lattice spacing is $a \simeq 0.11 \pm 0.01$ fm.
This gives:
\be
\lambda_A = 0.34 \pm 0.02 \pm 0.03 \, {\rm fm} ~~~~~~ (a \cdot m_q = 0.01) 
~.
\ee
The first error comes from our determination [Eq. (2.8)], the second from 
the error in converting the lattice spacing to physical units.
From $a \cdot m_q = 0.01$ to $a \cdot m_q = 0.02$ the value of the effective
$\Lambda_F$, defined by Eqs. (2.4) and (2.5), can change in principle.
Anyway, no published determination of $\Lambda_F$ (i.e., of the lattice 
spacing in physical units) exists for $a \cdot m_q = 0.02$. 
Some data on the pseudoscalar and vector meson masses, for quark masses $a 
\cdot m_q$ larger than 0.01, have been published in Ref. \cite{Laermann90}.
We have tried to extract $\Lambda_F$ (i.e., the lattice spacing) from those 
data by using the same procedure of Ref. \cite{Laermann93}.
We estimate that from $a \cdot m_q = 0.01$ to $a \cdot m_q = 0.02$ the 
effective mass--scale $\Lambda_F$ does not change appreciably within the 
errors. Assuming, as an indication, the same value of $\Lambda_F$ as for 
$a \cdot m_q = 0.01$, we then get:
\be
\lambda_A = 0.29 \pm 0.01 \pm 0.03 \, {\rm fm} ~~~~~~ (a \cdot m_q = 0.02)
~.
\ee
The values (3.2) and (3.3) must be compared with the {\it quenched} value
\cite{DiGiacomo92,myref97}
\be
\lambda_A = 0.22 \pm 0.01 \pm 0.02 \, {\rm fm} ~~~~~~ ({\rm YM ~ theory}) ~.
\ee
Here the value $\Lambda_L \simeq 4.9 \pm 0.5$ MeV has been assumed for the
pure--gauge $\Lambda$--parameter \cite{Michael88}.
The correlation length $\lambda_A$ increases when going from chiral to 
{\it quenched} QCD and this tendency is confirmed by the fact that 
$\lambda_A$ decreases by increasing the quark mass.
Of course, a precise determination of $\lambda_A$ should be done with more 
realistic values  for the quark masses.

We now come to the gluon condensate. Our lattice provides us with a 
regularized determination of the correlators. At small distances $x$ a 
Wilson {\it operator--product--expansion} (OPE) \cite{Wilson69} is 
expected to hold.
The regularized correlators will then mix to the identity operator {\bf 1},
to the renormalized local operators of dimension four,
${\alpha_s \over \pi}$:$G^a_{\mu\nu} G^a_{\mu\nu}$: and 
$m_f$:$\bar{q}_f q_f$: ($f = 1, \ldots, N_f$, $N_f$ being the number of 
quark flavours), and to operators of higher dimension:
\be
{1 \over 2\pi^2} {\cal D}_{\mu\nu,\mu\nu}(x) \mathop\sim_{x\to0}
C_{\bf 1}(x) \langle {\bf 1} \rangle + C_g(x) G_2 + 
\displaystyle\sum_{f=1}^{N_f} C_f(x) m_f \langle :\bar{q}_f q_f: \rangle
+ \ldots ~.
\ee
The mixing to the identity operator $C_{\bf 1}(x)$ shows up as a $c/|x|^4$ 
behaviour at small $x$. The mixings to the operators of dimension four 
$C_g(x)$ and $C_f(x)$ are expected to behave as constants for $x \to 0$, 
while the other Wilson coefficients in the OPE (3.5) are expected to vanish 
when $x \to 0$ (for dimensional reasons).
The coefficients of the Wilson expansion are usually determined in 
perturbation theory and are known to be plagued by the so--called 
{\it infrared renormalons} (see for example Ref. \cite{Mueller85} and 
references therein). In our case this means that, due to the infrared 
renormalon pole, terms coming from the mixing to the identity operator 
can produce by resummation a term which simulates a mixing to a condensate 
\cite{Mueller85}. There is no specific recipe to disentangle this 
``{\it perturbative}'' contribution to the condensates from a possible
``{\it genuine}'' value of them. A similar problem is present in any
Wilson OPE in QCD, and in particular in the expansion which leads to the
SVZ sum rules \cite{SVZ79}.

A practical way out, which provides good results for the sum rules, is to 
assume that the leading perturbative determination of the Wilson 
coefficients is a good approximation to the unknown determination which 
should be done by perturbing around the real vacuum of the theory (see for 
example Ref. \cite{Zakharov92} for a discussion about this point).
In this spirit, we shall assume that the renormalon ambiguity can be safely 
neglected in the extrapolation for $x \to 0$ of our correlators.
With the normalization of Eq. (3.5), this gives $C_g(0) \simeq 1$.
On the same line, the contribution from the quark operators in (3.5) can be 
neglected, because the corresponding condensates 
$m_f \langle :\bar{q}_f q_f: \rangle$ are much smaller than $G_2$ 
and the mixing coefficients $C_f(x)$ are of higher order than $C_g(x)$ 
in the perturbative expansion.
Within these approximations, we get the following expression for the gluon 
condensate, in terms of the parameters defined in Eq. (2.7):
\be
G_2 \simeq {6 \over \pi^2} (A_0 + A_1) ~.
\ee
At $a \cdot m_q = 0.01$ this gives, in physical units,
\be
G_2 = 0.015 \pm 0.003 \,^{+0.006}_{-0.003} \, {\rm GeV}^4 ~~~~~~ 
(a \cdot m_q = 0.01) ~.
\ee
At $a \cdot m_q = 0.02$ we obtain:
\be
G_2 = 0.031 \pm 0.005 \,^{+0.012}_{-0.007} \, {\rm GeV}^4 ~~~~~~ 
(a \cdot m_q = 0.02) ~.
\ee
The two values (3.7) and (3.8) should be compared with the corresponding value 
in the {\it quenched} theory \cite{myref97}:
\be
G_2 = 0.14 \pm 0.02 \,^{+0.06}_{-0.05}  \, {\rm GeV}^4 ~~~~~~ 
({\rm YM ~ theory}) ~.
\ee
The gluon condensate $G_2$ appears to increase with the quark mass, as 
expected, tending towards the asymptotic (pure--gauge) value of Eq. (3.9). 
Contrary to the previous discussion for $\lambda_A$, we 
have here a theoretical tool to understand the dependence of $G_2$ on the 
quark masses. According to Ref. \cite{NSVZ81}, we expect the following 
low--energy theorem to hold for small quark masses ($m_f \ll \mu$, $\mu$ 
being the renormalization scale):
\be
{d \over dm_f} \langle {\alpha_s \over \pi} :G^a_{\mu\nu}G^a_{\mu\nu}:
\rangle = -{24 \over b} \langle :\bar{q}_f q_f: \rangle ~,
\ee
where $b = 11 - {2 \over 3} N_f$, for a gauge group $SU(3)$ and $N_f$ quark 
flavours. 
For the two gluon condensates (3.7) and (3.8) one must use the renormalized 
quark masses $m_f$ \cite{quark-mass} corresponding to $a \cdot m_q = 0.01$ 
and $a \cdot m_q = 0.02$ respectively: for $a \cdot m_q = 0.01$ we have 
approximately $m_f \simeq 44$ MeV.
Making use of the popular values for the quark condensate ($\langle \bar{q} 
q \rangle \simeq -0.013 \, {\rm GeV}^4$ \cite{Laermann93,Dosch95}) and for the 
physical quark masses ($m_u \simeq 4$ MeV, $m_d \simeq 7$ MeV and $m_s \simeq 
150$ MeV), we can thus extrapolate from the value (3.7) to the {\it physical} 
gluon condensate, obtaining:
\be
G_2^{(physical)} \sim 0.022 \, {\rm GeV}^4 ~.
\ee
The procedure used is the same as in Ref. \cite{NSVZ81}.
The prediction (3.11) agrees with the empiric value obtained from experiments 
\cite{Dosch95,Narison96}: 
$G_2^{(empiric)} \simeq 0.024 \pm 0.011 \, {\rm GeV}^4$.

\bigskip
\noindent {\bf Acknowledgements}
\smallskip

The numerical simulations were done using the 512 node APE--QUADRICS in 
Pisa. We are particularly grateful to R. Tripiccione for help, and to G. Boyd
for having provided the HMC code and for his collaboration to the production 
of the configurations used in this work.

\vfill\eject

{\renewcommand{\Large}{\normalsize}
}

\vfill\eject

\noindent
\begin{center}
{\bf FIGURE CAPTIONS}
\end{center}
\vskip 0.5 cm
\begin{itemize}
\item [\bf Fig.~1.] The functions ${\cal D}_\perp^L f(\beta)^{-4}$ 
(upper curve) and ${\cal D}_\parallel^L f(\beta)^{-4}$ (lower curve)
versus physical distance, for quark mass $a \cdot m_q = 0.01$. 
The curves correspond to our best fit [Eqs. (2.7) and (2.8)].
\bigskip
\item [\bf Fig.~2.] The same as in Fig. 1 for quark mass $a \cdot m_q = 0.02$.
The curves correspond to our best fit [Eqs. (2.7) and (2.9)]. 
\end{itemize}

\vfill\eject

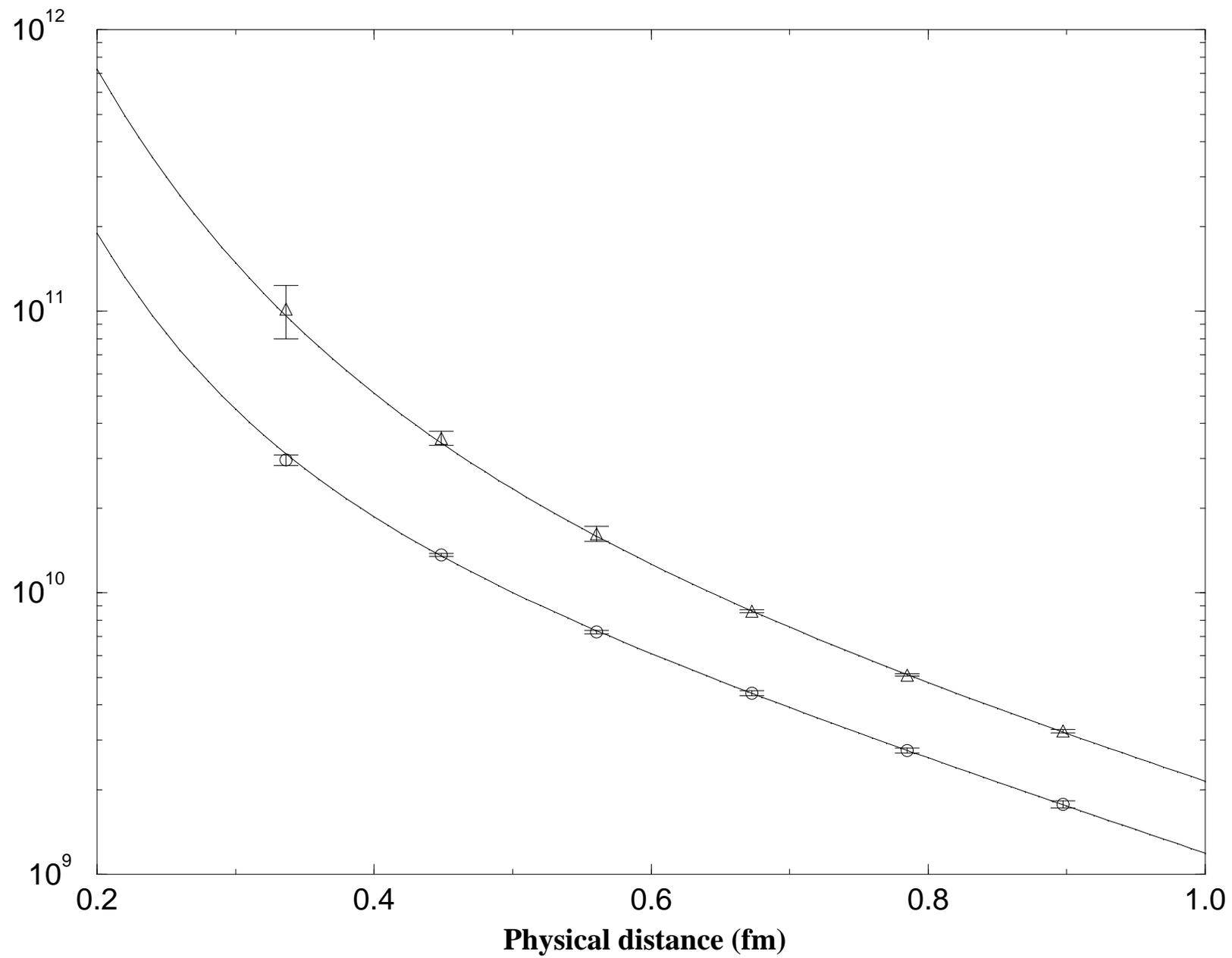
\end{document}